\documentclass[aps,twocolumn,groupedaddress,showpacs]{revtex4-1}
\usepackage{graphicx,amssymb,amsbsy,color}
\input epsf % defines \epsfbox and supporting macros
\epsfverbosetrue % messages will show height and width

\begin{document}

\title{Vortices in a rotating Bose-Einstein condensate under extreme elongation
in a harmonic plus quartic trap }
\author{C. C. Huang, C. H. Liu, and W. C. Wu}
\affiliation{Department of Physics, National Taiwan Normal
University, Taipei 11650, Taiwan}

\begin{abstract}
The behaviors of a rapid rotating Bose-Einstein condensate under
extreme elongation in a 2D anisotropic harmonic plus quartic trap
are investigated. Due to the quartic trap, the system remains
stable at high rotating velocity, $\Omega\geq \omega_\perp$
($\omega_\perp$ is the radial harmonic trap frequency), and vortex
lattices form even in the absence of the repulsive $s$-wave
interaction ($g$). When $g$ is present, the interplay between $g$
and the quartic trap potential can lead to rich vortex lattice
transition states as a function of $\Omega$, to which vortex
lattices vanish eventually at some higher $\Omega$.
\end{abstract}
\maketitle

\section{Introduction}
In recent years vortices in a rotating Bose-Einstein condensate
(BEC) have been extensively studied theoretically and
experimentally. Rotating condensates are typically confined in a
harmonic potential such that large vortex arrays are obtained when
angular velocity $\Omega$ is smaller than but approaching to the
radial trap oscillator frequency $\omega_\perp$. For instances in
Refs.~\cite{Matthews99,Madison00,Shaeer01,Raman01,Haljan01,Engels02,Schweikhard04},
a single vortex state or a state of several hundreds of vortices
have been successfully created. Theoretically, the behaviors of
the rapid rotating BEC systems are usually studied based on the
lowest Landau level (LLL) approximation
\cite{Ho01,Baym04,Cooper04,Watanabe04,Aftalion05,Watanabe06}. When
angular velocity is greater than the radial trap oscillator
frequency, $\Omega>\omega_\perp$, the system becomes unstable due
to the ineffectiveness of confinement.

In the experiment of Bretin {\em et. al} \cite{Bretin04}, a
quartic trapping potential has been successfully created. Several
groups then consider the type of trap potential $V(x,y) =
\frac{1}{2}m{\omega_\perp^2} {r_\perp^2} +
\frac{1}{4}u{r_\perp^4}$, where $m$ is the atom mass, $u$ is the
strength of the quartic potential, and $r_\perp^2=x^2+y^2$
\cite{Blanc08,Jackson04,Jackson04-1,Fetter01,Fetter05}. With this
quartic potential, due to the effect of strong confinement, one
thus can study the rotating Bose condensates at higher angular
velocities, $\Omega\geq\omega_\perp$. It is understood that the
centrifugal force, varying as $\Omega^2 r_\perp$, can always be
compensated by the trapping force varying as
$-(m\omega_\perp^2r_\perp+ur_\perp^3)$.

Rotating BECs have also been studied under an anisotropic harmonic
trap for a quasi-two-dimensional (2D) system
\cite{Oktel04,Fetter07}. Moreover, by manipulating the anisotropic
trap frequencies associated with the rotating velocity, quasi-2D
rotating BEC under the anisotropic harmonic trap can be reduced to
the limit of extreme elongation \cite{Sinha05,Lotero05,Matveenko09}.
In Ref.~\cite{Sinha05} the authors have studied the roton-maxon
excitation spectrum in a rapid rotating weakly interacting BEC
under extreme elongation in a 2D anisotropic harmonic trap.
When the interaction or the angular velocity increases,
the system could undergo a second-order quantum transition to a state of a
periodic structure -- rows of vortices \cite{Sinha05}. Similar
transitions have also been investigated in Refs.~\cite{Lotero05,Matveenko09}.
A common finding is that the larger the interaction or angular velocity is,
the more the vortex lattice row number is.

Inspired by the works
mentioned above \cite{Bretin04,Sinha05,Lotero05,Matveenko09}, it is
of particular interest to study how the rapid rotating BEC behaves
under the extreme elongation if a quartic trapping potential is
added.  With this goal, this paper considers the following effective
2D trap potential
\begin{eqnarray}
V(x,y)={1\over 2}m[\omega_x^2x^2+\omega_y^2y^2]+{1\over 4}ux^4,
\label{eq:mainpotential}
\end{eqnarray}
where, with the strength $u$, the anharmonic quartic potential acts
on the $x$-axis only. By adding the quartic potential along the $x$
axis only, the effect of the anharmonic trap for the extreme
elongated system can be seen in a more transparent way (effective 1D
problem). In fact, the kind of effective trap
(\ref{eq:mainpotential}) can be realized in a realistic experiment.
One can consider a quasi-2D system that is confined in the $xy$ plane
by a harmonic trap of relatively large $\omega_z$. Now, consider
that an extra isotropic 2D quartic trap, $V_4\sim (x^2+z^2)^2$, is
added and suppose that the rotating angular frequency is purely along the
$z$ direction. Since the system is already confined in the $xy$
plane, any quartic trap effect in the $z$ direction will simply
enhance the confinement of the system in the $xy$ plane. This means
effectively, $V_4\sim x^4$.

With the quartic trapping potential (\ref{eq:mainpotential}) which is
introduced to stabilize the system against decay at higher rotation
velocities ($\Omega$), one can then study the system even for
$\Omega\geq\omega_x$. It will be shown that in the limit of no
interaction, ground state forms no vortex lattice at smaller $\Omega$ and
when $\Omega$ becomes larger, one row of vortex lattice will form.
Vortex lattice will melt eventually
at some larger $\Omega$. When the interaction is turned on,
melting of the vortex lattice at some higher $\Omega$
is also found. Nevertheless, rich
transition states occur for the vortex lattices. The transition states,
arising due to the instability at a critical interaction strength or
a critical rotation velocity,
were first studied in Ref.~\cite{Sinha05} in an
elongated system without the quartic potential. One will see
that a similar instability leading to the
formation of a periodic structure of vortex rows occurs in the current context.
Of most interest, a parameter labeled by $N_c$
which corresponds to number of terms involved in the $k$-sum of the
wavefunction is intimately related to the
number of vortex rows \cite{Sinha05}.

The paper is organized as follows. In Sec.~\ref{sec2}, we outline
the GP energy functional essential for a rapid rotating BEC
system under extreme elongation in a 2D anisotropic harmonic plus quartic trap.
Details about how the ground-state wave function is
obtained and how the total energy is minimized are given. In
Sec.~\ref{sec3}, we study the properties of the system without the
interaction. This section is devoted to fully understand the
effect of the quartic potential. It is shown that the behaviors of
the system with a quartic potential are quite different compared
to those of no quartic potential. In Sec.~\ref{sec4}, effect of
the $s$-wave interaction $g$ is studied, which is treated by the
perturbation method. It will be shown that the interplay between
$g$ and the quartic trap potential can lead to rich vortex lattice
transition states as a function of $\Omega$, to which vortex
lattices vanish eventually at some higher $\Omega$.
Sec.~\ref{sec5} is a brief conclusion.

\section{GP energy functional}\label{sec2}
The GP energy functional of a rapid rotating, quasi-2D BEC system
can be given by
\begin{eqnarray}
E = \int {dxdy \Psi_0^*\left[-\frac{{{\hbar ^2\nabla ^2}}}{{2m}} +
V(x,y) + \frac{g_{\rm 2D}}{2}{{\left| {{\Psi_0}} \right|}^2} - {\bf
\Omega}\cdot{\bf L}\right]\Psi_0}, \nonumber\\
\label{eq:orgine}
\end{eqnarray}
where the trap potential $V(x,y)$ was given in
Eq.~(\ref{eq:mainpotential}), $\Psi_0=\Psi_0(x,y)$ is the wave
function, $g_{\rm 2D}\equiv 4\pi\hbar^2a_s/(ma_z\sqrt{2\pi})$ is the
effective 2D $s$-wave interaction with $a_s$ the $s$-wave scattering
length and $a_z\equiv\sqrt{\hbar/m\omega_z}$
associated with the harmonic trap frequency $\omega_z$ \cite{Petrov00},
${\bf\Omega}$ is the rotation angular velocity, and ${\bf L}$ is the
angular momentum. When ${\bf\Omega}=\Omega\hat{z}$, the rotating
term becomes
\begin{eqnarray}
- \Omega \Psi_0^*{L_z}{\Psi_0} = i\hbar \Omega \Psi_0^*\left[
{x\frac{\partial }{{\partial y}} - y\frac{\partial }{{\partial
x}}} \right]{\Psi_0}. \label{eq:rotation}
\end{eqnarray}
Eq.~(\ref{eq:orgine}) can also be written as
\begin{eqnarray}
E &=& \int dxdy \Psi_0^*\label{eq:newenergy}\\
&\times&  \left[ {\frac{{{{(- i\hbar\nabla + m {\bf r}_\perp \times
{\bf \Omega})}^2}}}{{2m}} + V_{\rm eff}(x,y) +
\frac{g_{\rm 2D}}{2}{{\left| {{\Psi_0}} \right|}^2}}
\right]{\Psi_0},\nonumber
\end{eqnarray}
where ${\bf r}_\perp=(x,y)$ and $V_{\rm
eff}(x,y)\equiv(m/2)[(\omega_x^2-\Omega^2)x^2+
(\omega_y^2-\Omega^2)y^2]+(u/4)x^4$. For the part of the
anisotropic harmonic trap, trap frequencies can be written as
$\omega_x^2\equiv \omega_0^2+\Lambda^2$ and
$\omega_y^2\equiv\omega_0^2-\Lambda^2$. If one sets
$\Lambda^2=\omega_0^2-\Omega^2$, the effective potential becomes
\begin{eqnarray}
V_{\rm eff}(x,y) =m(\omega_0^2-\Omega^2)x^2+{1\over 4}ux^4,
\label{eq:effp}
\end{eqnarray}
leaving no $y$ dependence in it. This means that the system will
reach a state under extreme elongation along the $y$ direction. This
is the case of most interest in the current context (effective 1D
case). Nevertheless, for practical reason, we will assume that the
system is confined in a large rectangular box of side length $L$.
When $u=0$ (no quartic trap), as mentioned before, the system will
become unstable for $\Omega \geq\omega_0$ unless the $s$-wave
interaction is attractive ($g_{\rm 2D}<0$). However when $u\neq0$, the
system remains stabilized for $\Omega \geq\omega_0$ and $g_{\rm 2D}>0$.

To solve the wavefunction of the system, we shall use an approach
similar to that used in Refs.~\cite{Sinha05,Lotero05}. Firstly, we
apply the gauge transformation, ${\Psi_0} = \Psi e^{-i m\Omega
xy/\hbar}$, and consequently the energy
functional~(\ref{eq:newenergy}) becomes
\begin{eqnarray}
E &=&\int dxdy {\Psi ^*}\label{eq:fine}\\
&\times&  \left[ -{ \frac{{{\hbar ^2}{\nabla ^2}}}{{2m}}+
 {V}_{\rm eff}^\prime(x,y) + \frac{g_{\rm 2D}}{2}{{\left| \Psi  \right|}^2}+
 2i\hbar \Omega x\frac{\partial }{{\partial y}}} \right]\Psi \nonumber
\end{eqnarray}
with ${V}_{\rm eff}^\prime(x,y)=m(\omega_0^2+\Omega^2)x^2+(u/4)x^4$.
Secondly the $s$-wave interaction $g_{\rm 2D}$ is assumed to be weak
such that wavefunctions corresponding to no interaction case
($g_{\rm 2D}=0$) can be used as the basis functions. Thirdly, the effect
of the $s$-wave interaction will be reinstalled and treated by the
perturbation method. Thus it is proposed that the wave function
$\Psi$ is a linear combination of the lowest level eigenfunction(s),
$\phi_k$, associated with energy functional (\ref{eq:fine}) with
$g_{\rm 2D}=0$. That is,
\begin{eqnarray}
\Psi  = \sqrt{N}\sum\limits_k {{C_k}{\phi _k}}, \label{eq:orgfun}
\end{eqnarray}
where $N$ is the total number of particles and it requires that
$\sum_k {{\left|{{C_k}}\right|}^2}=1$ for completeness. Here
$k\equiv 2\pi \ell/L$ with $\ell$ an integer. Since $L$ is large,
$k$ can be treated in the continuous limit. When the quartic trap
vanishes ($u=0$), $\phi_k$ corresponds to LLL. Furthermore,
eigenfunctions $\phi_k$ are assumed to take the separate form
\begin{eqnarray}
{\phi _k} = A{e^{iky}}{\chi _k}\left( x \right),
\label{eq:funform}
\end{eqnarray}
where $A=1/\sqrt{L}$ is the normalization constant. Substituting
Eqs.~(\ref{eq:orgfun}) and (\ref{eq:funform}) into
Eq.~(\ref{eq:fine}) with $g=0$, one obtains the one-dimensional
(dimensionless) differential equation for ${\chi _k}\left( x
\right)$,
\begin{eqnarray}
\left[ { - \frac{{{d^2}}}{{d{x^2}}} + \left( {2 +
\frac{{{2\Omega ^2}}}{{\omega _0^2}}} \right){x^2} -
\frac{{4\Omega }}{{{\omega _0}}}kx + \lambda {x^4} + {k^2}}
\right]{\chi _k} \nonumber\\
= {\varepsilon _k}{\chi _k}, \label{eq:onedim}
\end{eqnarray}
where ${\varepsilon _k}$ is the eigenenergy. Here $\lambda\equiv
u\hbar/2m^2\omega_0^3$ and all the lengths and energies are scaled
by the unit length $a_0\equiv\sqrt{\hbar/m\omega_0}$ and unit
energy $E_0\equiv\hbar\omega_0/2$. When $\lambda=0$, one has
analytic solutions for Eq.~(\ref{eq:onedim}) \cite{Lotero05}. But
for $\lambda\neq 0$, no analytic solution is available and
numerical computation is needed. Here we shall apply the finite
difference method to find solutions of the normalized function
$\chi_k$.

With Eqs.~(\ref{eq:orgfun})--(\ref{eq:onedim}), the full GP energy
functional (\ref{eq:fine}) becomes
\begin{eqnarray}
 \frac{E}{N} &=& \sum\limits_k^{N_c}{{\left| {{C_k}} \right|}^2}
 {\varepsilon _k} + \frac{ng}{2} \sum \limits_{{k_1},{k_2},{k_3},{k_4}}^{N_c}
 {C_{{k_1}}^*C_{{k_2}}^*{C_{{k_3}}}{C_{{k_4}}}} \nonumber \\
 &\times& \int {dx{\chi _{{k_1}}}{\chi _{{k_2}}}{\chi _{{k_3}}}
 {\chi _{{k_4}}}{\delta _{{k_3} + {k_4},{k_1} + {k_2}}}},
  \label{eq:E}
\end{eqnarray}
where $n\equiv N/(L/a_0)$, $g\equiv 4\sqrt{2\pi}a_s/a_z$ represents the
new scale of $g_{\rm 2D}$, and $N_c$ is the number of terms used
in the linear combination. $N_c$ is determined upon the condition
that energy functional is minimized. The first part of the GP
functional~(\ref{eq:E}) involves the kinetic, harmonic, anharmonic
(quartic), and rotational energies. While the second part
corresponds to the interaction energy obtained by the perturbation
manner.  It is worth noting in Eq.~(\ref{eq:onedim}) that the
function $\chi_k(x)=\chi_{-k}(-x)$ and the energy
$\varepsilon_k=\varepsilon_{-k}$. Due to the symmetry of
$\varepsilon_k=\varepsilon_{-k}$, it is useful to write
Eq.~(\ref{eq:orgfun}) as
\begin{eqnarray}
\Psi  = \sqrt N \left[ {{C_0}{\phi _0} + \sum\limits_{i = 1}^j
{({C_{{k_i}}}{\phi _{{k_i}}} + {C_{ - {k_i}}}{\phi _{ - {k_i}}})}
} \right].
 \label{eq:finfun}
\end{eqnarray}
With Eq.~(\ref{eq:finfun}), it's evident that there are two cases
for $N_c$. One is $N_c=2j+1$ if $C_0\neq 0$ and another is
$N_c=2j$ if $C_0=0$.

\begin{figure}[tb]
\vspace{0.0cm}
\includegraphics[width=0.52\textwidth]{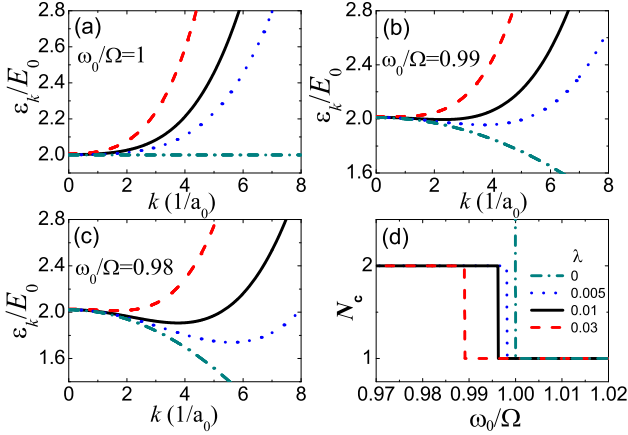}
\vspace{-0.5cm} \caption {(Color online) In frame (a)-(c), lowest eigenenergy of
Eq.~(\ref{eq:onedim}), $\varepsilon_k$, is plotted as the function
of $k$ for three cases: $\omega_0/\Omega=1.0, 0.99$ and $0.98$
respectively. Frame (d) plots $N_c$ (see text) as the function of
$\omega_0/\Omega$. In each frame, the four lines correspond to
$\lambda=0.0$, $0.005$, $0.01$, and $0.03$ respectively.
For the units, $E_0=\hbar\omega_0/2$ and
$a_0=\sqrt{\hbar/m\omega_0}$.}
\label{fig1}
\end{figure}

\section{Noninteracting system}\label{sec3}

In this section, we consider the ground-state properties of a
noninteracting system ($g=0$). It will be shown for this
noninteracting system that when $\lambda\neq 0$, the properties of
the system are quite different from those of the $\lambda= 0$
case. It is useful to first review the results for the $\lambda=
0$ case \cite{Lotero05}. When $\lambda= 0$, one has the analytic
solution for energy $\varepsilon_k$ in Eq.~(\ref{eq:onedim})
\begin{eqnarray}
\varepsilon_k\rightarrow {\varepsilon^0_k} = \sqrt {2(1 +
\frac{{{\Omega ^2}}}{{\omega _0^2}})}  + {k^2}\left( {\frac{{1 -
\frac{{{\Omega ^2}}}{{\omega _0^2}}}}{{1 + \frac{{{\Omega
^2}}}{{\omega _0^2}}}}} \right).
 \label{eq:analye}
\end{eqnarray}
The results of Eq.~(\ref{eq:analye}) can be divided into three
folds. (i) When $\Omega<\omega_0$, minimum $\varepsilon^0_k$
occurs at $k=0$. Consequently the ground-state wavefunction of the
system is just $\Psi=\sqrt{N}C_0\phi_0$ ($N_c=1$). (ii) When
$\Omega=\omega_0$, $\varepsilon^0_k$ is the same for all $k$'s. In
this highly degenerate case, the ground-state wavefunction should
include all eigenstates ($N_c=\infty$). (iii) When
$\Omega>\omega_0$, minimum $\varepsilon^0_k$ occurs at
$k\rightarrow\infty$. It indicates that the system is unstable. An
important consequence of the $g=\lambda=0$ case is that vortex
(lattice) will not form regardless of the ratio of
$\omega_0/\Omega$.

When $\lambda\neq 0$, in contrast, behaviors of the system can be
very different. We first consider the case when $\lambda$ is small
such that wavefunction of the $\lambda=0$ case can be used to
obtain the $\lambda$ correction perturbatively. The leading
correction of the energy is found to be ($\varepsilon_k \simeq
\varepsilon^0_k +\Delta\varepsilon^0_k$)
\begin{eqnarray}
{\Delta\varepsilon^0_k}=\lambda({A_0}+{A_2}{k^2}+{A_4}{k^4}),
 \label{eq:pere}
\end{eqnarray}
where $A_0=3/(8R)$, $A_2=3\Omega^2/(\sqrt{2}\omega_0^2R^{5/2})$,
and $A_4={\Omega^4}/{(\omega_0^4 R^4)}$ with $R\equiv
1+\Omega^2/\omega_0^2$. Note that all three $A_i$ are positive, so
the correction $\Delta\varepsilon^0_k>0$. Moreover, it is found
that the behaviors of the wavefunction $\Psi$ can be divided into
two branches. When $\Omega\leq(1+3\lambda/8)\omega_0$, minimum
$\varepsilon_k$ occurs at $k=0$. Consequently
$\Psi=\sqrt{N}C_0\phi_0$ ($N_c=1$). In contrast when
$\Omega>(1+3\lambda/8)\omega_0$,
$\Psi=\sqrt{N}(C_{k_1}\phi_{k_1}+C_{-k_1}\phi_{-k_1})$ ($N_c=2$),
where $k_1$ corresponds to where minimum $\varepsilon_k$ is.
Comparing these small but finite $\lambda$ results to those of
$\lambda=0$ discussed before, one sees clearly that the general
behaviors are quite different between the $\lambda=0$ and
$\lambda\neq 0$ cases, especially in the regime
$\Omega\agt\omega_0$.

\begin{figure}[tb]
\vspace{0.0cm}
\includegraphics[width=0.45\textwidth]{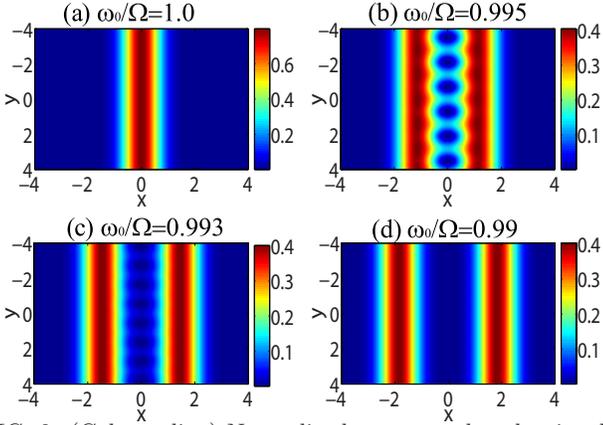}
\vspace{-0.5cm} \caption {(Color online) Normalized atom number density
distribution, $|\Psi|^2/n$, is plotted in the $xy$ plane for (a)
$\omega_0/\Omega =1.0$, (b) 0.995, (c) 0.993, and (d) 0.99. Here
$\lambda$ is 0.005 for all frames. As shown, single vortex line
exhibits in frame (b) and (c) only. $x$ and $y$ axes are in units of
$a_0=\sqrt{\hbar/m\omega_0}$.}
 \label{fig2}
\end{figure}

When $\lambda$ is not too small such that perturbation approach is
no longer valid, one needs to solve Eq.~(\ref{eq:onedim})
numerically for $\varepsilon_k$. In Fig.~\ref{fig1}(a)--(c) with
 $\omega_0/\Omega=1.0$, $0.99$, and $0.98$
respectively, we numerically solve and plot $\varepsilon_k$ as the
function of $k$. Three finite $\lambda$ cases ($\lambda=0.005$,
$0.01$, and $0.03$) are considered. The $\lambda=0$ case is included
for comparison. Based on the experimental data given in
Ref.~\cite{Bretin04} that $w_x\sim 2\pi\times64.8$ Hz corresponds to
$\lambda\simeq 0.001$, it is estimated that $w_x\sim 2\pi\times38.5$,
$2\pi\times30$, and $2\pi\times21$ Hz will correspond to
$\lambda=0.005$, $0.01$, and $0.03$ cases respectively.
As shown in Fig.~\ref{fig1}(a)
for $\omega_0/\Omega =1.0$, $k_1$ (corresponding to minimum
$\varepsilon_k$) is zero for all three finite $\lambda$ cases. This is
in big contrast to the $\lambda=0$ case where $\varepsilon_k^0$ is
$k$ independent [see Eq.~(\ref{eq:analye})].
In the case of $\omega_0/\Omega =0.99$, $k_1$
is finite for $\lambda=0.01$ and $0.005$ cases but is zero for the
$\lambda=0.03$ case [see Fig.~\ref{fig1}(b)]. While the corresponding
$\varepsilon_k^0$ becomes lesser when $k$ is larger and no minimum-energy
state is found. Finally for the case of $\omega_0/\Omega =0.98$
[see Fig.~\ref{fig1}(c)],
$k_1$ is finite for three $\lambda\neq0$ cases. Similar to the
$\omega_0/\Omega =0.99$ case, $\varepsilon_k^0$
will become lesser when $k$ is larger and no minimum-energy
state is found.
Thus for the larger $\lambda$ ($=0.03$) case, vortex forms only
when the angular velocity ($\Omega$) is larger.
%And behavior of the energy
%$\varepsilon_k$ also becomes more sharp in small $k$.
Moreover with the same $\lambda$, the value of $k_1$ (corresponding
to minimum $\varepsilon_k$) is larger when $\Omega$ is larger. This
means that the number of vortices per unit length (in the elongated
direction) will be larger when $\Omega$ is larger. In
Fig.~\ref{fig1}(d),  $N_c$ is plotted as the function of
$\omega_0/\Omega$ for different $\lambda$ cases.

In practice, it is useful to compare the  energy scale of the
(dimensionless) quartic coupling $\lambda$ to the harmonic trap
frequency $\omega_x$ or $\omega_y$. When $\Omega\sim\omega_0$,
$\omega_x^2/\omega_0^2=2-(\omega_0/\Omega)^2\sim 1$. Since the
results discussed above are for $\lambda\alt 0.03$, the effect of
the quartic trap is seen to be quite drastic.

In Fig.~\ref{fig2}, normalized atom number density distributions,
$|\Psi|^2/n$, are plotted in the $xy$ plane. Four cases are
considered, namely $\omega_0/\Omega=1.0$, $0.995$, $0.993$, and
$0.99$ with $\lambda=0.005$ in all cases. In the case of
$\omega_0/\Omega=1.0$ [Fig.~\ref{fig2}(a)], no vortex is formed.
When the rotation velocity $\Omega$ is increased, single vortex
line is seen to occur for both the $\omega_0/\Omega=0.995$ and
$0.993$ cases [Fig.~\ref{fig2}(b) and (c)]. It is noted, however,
that number of vortices per unit length is larger for the
$\omega_0/\Omega=0.993$ case as compared to that of the
$\omega_0/\Omega=0.995$ case.
%Although the vortex lattice is less
%evident in the former case.
When $\Omega$ is further increased to above a critical value,
$\omega/\Omega_c\alt 0.99$, vortex state disappears (melts) due to
the large centrifugal force [see Fig~\ref{fig2}(d)]. At this large
$\Omega$ case, atoms are pushed to two sides along the elongated
potential well and no atoms are left in the center.

The most important results obtained in this section for the
noninteracting system ($g=0$) are summarized as the following. (i)
The quartic trap can make the system remain stable at higher
rotation velocity ($\Omega>\omega_0$). (ii) Single vortex line can
exhibit at some $\Omega$ regimes. (iii) The single-line vortex
lattice will vanish eventually at some higher $\Omega$ to which
atoms are push to two sides along the elongated potential well. In
next section, the effect of interaction ($g\neq 0$) will be
discussed in details.

\section{Effect of interaction}\label{sec4}
\begin{figure}[tb]
\vspace{-0.5cm}
\includegraphics[width=0.50\textwidth,height=8cm]{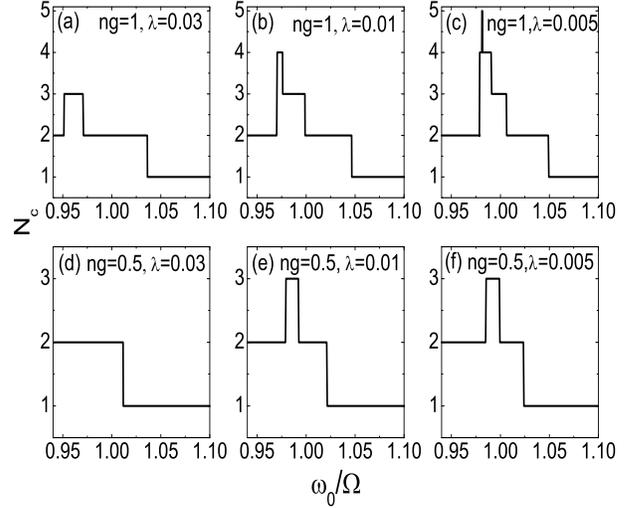}
\vspace{-1.1cm} \caption {$N_c$ is plotted as the
function of $\omega_0/\Omega$ for six cases: (a)
$(ng,\lambda)=(1,0.03)$, (b) (1,0.01), (c) (1,0.005), (d)
(0.5,0.03), (e) (0.5,0.01) and (f) $(0.5,0.005)$. Transition states
occur in all frames except (d).} \label{fig3}
\end{figure}

In this section, the effect of interaction ($g$) is studied by
the perturbation approach. Similar to previous section, we consider
also the three finite $\lambda$ cases:
$\lambda=$ $0.03$, $0.01$, and $0.005$. When
the energy associated with the quartic trap is smaller than the energy
associated with the harmonic trap and when $\Omega\sim\omega_0$, the energy
difference between lowest and first excited band of $\varepsilon_k$ is about
$2\hbar\omega_0$. In our dimensionless scale, $2\hbar\omega_0=4$.
Thus the perturbative approach is valid as long as
\begin{eqnarray}
\frac{g}{2}{{\left| \Psi \right|}^2}\ll 4 ~~~{\rm or}~~~ ng\left(
\frac{{\left|\Psi \right|}^2}{n}\right)\ll 8.
 \label{eq:condition}
\end{eqnarray}
The value of ${\left| \Psi \right|}^2/n$ is roughly $0.4$ (see
Figs.~\ref{fig2} and \ref{fig4}). Thus when
$ng\ll 20$, the perturbation approach is valid.
%For example the authors use the $ng=1.91$($<\frac{1}{10}\times20$)
%in the paper\cite{Lotero05}.
Hence the values of $ng=0.5$ and
$1.0$ considered here should be safely valid for the perturbative approach.
Typically $n$ is
more than $1$ in a 2D system \cite{Sinha05}. If
$N=10^4$ and $L/a_0=100$, the value of $n$ is about $100$. For
$ng=1$, it corresponds to $a_s/a_z\simeq10^{-3}$. In
Fig.~\ref{fig3}, the number of $N_c$ [terms involved in the
summation (\ref{eq:E})] is determined and plotted as the function of
$\omega_0/\Omega$ for six cases, namely $(ng,\lambda)=(1,0.03)$,
$(1,0.01)$, $(1,0.005)$, $(0.5,0.03)$, $(0.5,0.01)$, and
$(0.5,0.005)$. It turns out that the results of $N_c$ are quite
fascinating in terms of the change of $\Omega$ and the interplay
between the strength of the interaction and the strength of the
quartic potential. One common feature of all frames in
Fig.~\ref{fig3} is that $N_c$ shifts from 1 (at lower $\Omega$) to 2
eventually (at higher $\Omega$). In between of the 1 to 2 period,
transition states occur for a large $\Omega$ span. For example,
$N_c\geq 3$ transition states can occur in all frames except in
frame (d). Besides, $N_c$ increases in a trend as $\Omega$ increases
for the transition states. The reason why $N_c$ all shift from 1 (at
lower $\Omega$) to 2 eventually (at higher $\Omega$) can be
understood as follows. One recalls the $N_c$ results for $g=0$ in
Fig.~\ref{fig1}(d), where all three finite $\lambda$ cases are observed
to exhibit the $N_c=1\rightarrow 2$ transition (at different
critical $\Omega$ though).

\begin{figure}[tb]
\vspace{-0.0cm}
\includegraphics[width=0.50\textwidth,height=4.5cm]{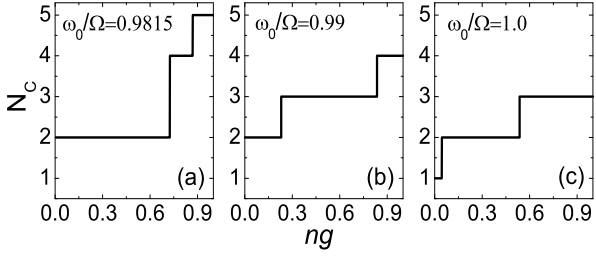}
\vspace{-1.1cm} \caption {$N_c$ is plotted as the
function of $ng$ for three cases: (a)
$\omega_0/\Omega=0.9815$, (b) 0.99, and (c) 1.0.
$\lambda=0.005$ in all three frames.} \label{fig3-1}
\end{figure}

%For the same $\lambda$ case,
%minimum $\varepsilon_{k_1}$ is lower for the larger $\Omega$ case.
When the interaction $g$ is turned on and still valid in the
perturbative regime, $g$ will only play little role in the large
$\Omega$ limit. It means that when $\Omega$ is large enough,
transition states, which arise due to the effect of $g$, will
disappear. As an extreme example shown in Fig.~\ref{fig3}(d),
because $g$ is relatively small ($ng=0.5$) and the quartic
coupling is relatively large ($\lambda=0.03$), consequently no
transition state occurs. When $\lambda$ is reduced [see
Fig.~\ref{fig3}(e) and (f)], or when $g$ is increased [see
Fig.~\ref{fig3}(a)], transition states will occur.

Moreover, it is found that the critical value of $\Omega$ to which
$N_c$ changes from 1 to 2 at the lower $\Omega$ side is smaller when
$\lambda$ is smaller (if $ng$ is fixed) or when $ng$ is larger (if
$\lambda$ is fixed). It implies that the system will enter the
vortex state earlier if $g$ is relatively larger or $\lambda$ is
relatively smaller. Furthermore, with the same $\lambda$, transition
states will sustain for a larger span of $\Omega$ if $g$ is larger
and can go up to a higher value of $N_c$. The latter means that
vortex lattice can have a higher row number. One sees in
Fig.~\ref{fig3}(c) that $N_c$ can go up to 5, although for a small
period of $\Omega$. Fig.~\ref{fig3-1} plots number $N_c$ as
the function of $ng$ with $\lambda=0.005$ and
$\omega_0/\Omega=0.9815$, $0.99$, and $1.0$ respectively.
Basically Figs.~\ref{fig3} and \ref{fig3-1} show how the
different regimes behave as the change of the interaction,
quartic trap strength, and rotation velocity.

\begin{figure}[tb]
\vspace{0.0cm}
\includegraphics[width=0.45\textwidth,height=10cm]{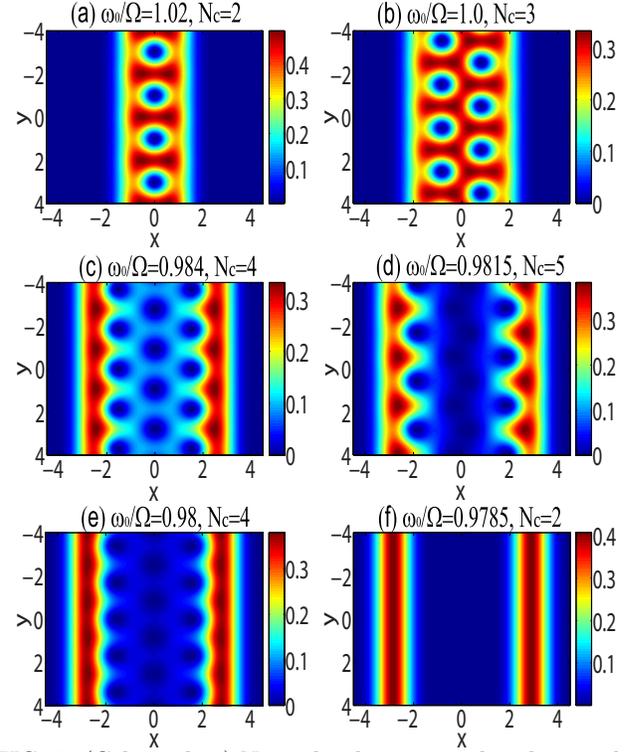}
\vspace{-0.5cm} \caption {(Color online) Normalized atom number density
distribution, $|\Psi|^2/n$, is plotted in the $xy$ plane for (a)
$\omega_0/\Omega =1.02$, (b) 1.0, (c) 0.984, (d) 0.9815, (e) 0.98,
and (f) 0.9785. Here $ng=1$ and $\lambda=0.005$ for all frames
[corresponding to the case in Fig.~\ref{fig3}(c)]. Vortex lattices
with the transitions of row number and lattice constant are observed
as $\Omega$ changes. $x$ and $y$ axes are in units of
$a_0=\sqrt{\hbar/m\omega_0}$.} \label{fig4}
\end{figure}

The normalized density profile in the $xy$ plane, $|\Psi|^2/n$, is
shown in Fig.~\ref{fig4} for $(ng,\lambda)=(1,0.005)$
[corresponding to the case in Fig.~\ref{fig3}(c)]. As seen in
Fig.~\ref{fig3}(c), the change of $N_c$ in the transition states
is quite rich for this case. Six angular velocities are studied,
namely $\omega_0/\Omega=1.02$, 1.0, 0.984, 0.9815, 0.98 and 0.9785
respectively for Fig.~\ref{fig4}(a)--(f). With these values of
$\Omega$, $N_c$ corresponds to 2, 3, 4, 5, 4, and 2 respectively.
One sees in Fig.~\ref{fig4}(a)--(d) that when $\Omega$ is
increased from $\omega_0/\Omega=1.02$ to $0.9815$, atoms are
pushed to the two sides and the number of vortices becomes more
and more. The row number of vortex line also increases from 1 to
4. While in Fig.~\ref{fig4}(e), the vortex row number is reduced
to 3 ($N_c=4$) again. In the case of Fig.~\ref{fig4}(f), although
$N_c$ is reduced to 2, but vortex lattice vanishes (melts) due to
the large centrifugal force. Similar vortex lattice melting
transition (atoms are completely pushed to the two sides) at large
$\Omega$ has already been seen in the previous section of no
interaction.
%Nevertheless, the rich behaviors of the transition
%states for the weakly-interacting system is not accessible in the
%non-interacting interaction system [see Fig.~\ref{fig2}].

\section{Conclusions} \label{sec5}

This paper investigates the effect of a quartic potential on a
fast rotating BEC system under the extreme elongation. In contrast
to the harmonic trap alone case where system is unstable when the
angular velocity $\Omega$ is larger than the radial trap
oscillator frequency $\omega_0$, the quartic trap can lead the
system to remain stable at higher rotation velocity
($\Omega>\omega_0$). The interplay between the weak $s$-wave
interaction and the quartic trap potential can result rich vortex
lattice transition states as a function of $\Omega$. At large
$\Omega$, atoms are eventually push to the two sides along the
elongated potential well.

\acknowledgements This work is supported by the National Science
Council, Taiwan under the Grant No. 96-2112-M-003-008. We also
acknowledge the support from NCTS, Taiwan.

%\bibliography{vortex}
%\bibliographystyle{prsty}

\end{document}